\documentclass[preprint,authoryear,12pt]{elsarticle}
\usepackage{amssymb}
\usepackage{graphicx}
\usepackage{epsfig}
\usepackage{amsthm}
\usepackage{longtable}
\usepackage{lscape}

\journal{????}

\begin{document}

\begin{frontmatter}

\title{IC 361:- Near-infrared and $UBVRI$ Photometric Analysis}


\author[ku,de]{Gireesh C. Joshi}
\ead{gchandra.2012@rediffmail.com}
\address[ku]{Department of Physics, Kumaun University, Nainital-263002}
\address[de]{Department of Physics, Govt. Degree College, Kotdwar-Bhabar, Pauri, Uttarakhand-246149 (INDIA)}
\def\astrobj#1{#1}
\begin{abstract}
The detailed optical and infra-red photometric analysis of the open star cluster IC\,361 have shown in the present manuscript. On studying the radial density profile, radial extent of the cluster is found to be $8.0\pm0.5$ arcmin. The basic physical parameters of the cluster such as $E(B-V)$ = $0.56\pm0.10$ mag, $E(V-K)=1.72{\pm}0.12$ mag, log(Age)=9.10$\pm$0.05, and $(m-M)_{0} = 12.54\pm0.05$ mag are obtained using the colour-colour and colour-magnitude diagrams. IC\,361 is found to be located at a distance of $3.22\pm0.07$ kpc. Using the archival proper motion catalogues, we are estimated the mean proper motions of IC\,361 as 4.97$\pm$0.17 mas~yr$^{-1}$ and -5.80$\pm$0.18 mas~yr$^{-1}$ in the direction of RA and DEC, respectively. We have not found steeper slop of stellar mass functions of cluster's main sequence. However, the mass function slope of coronal region is found to be $-1.06\pm0.09$ which is too low compare than Salpeter value. Our study is further showing dynamically relax behaviour of the cluster IC\, 361. 
\end{abstract}

\begin{keyword}
(Galaxy): Open star cluster; individual:  IC 361; variable: pulsation; method-data analysis

\end{keyword}

\end{frontmatter}


\def\astrobj#1{#1}
\section{Introduction}
\label{sec:intro}
IC 361 is an open cluster (OC) in the constellation Camelopardus \citep{pi78}. This cluster is well detached from the field, but it is very poorly studied due to its faintness \citep{zd09}. IC 361 lies in the second Galactic quadrant, in the immediate vicinity of the Camelopardalis dark clouds; as a result, it demonstrates a considerable interstellar reddening \citep{zd10}.\cite{pi78} were adopted a distance of 2.5 kpc, a high value of reddening, $E_{B-V} = 0.55$, and suggested the age in the range 0.5 to 1 Gyr. On this background, the comprehensive analysis of $UBVRI$ and $JHK$ photometric data is carried out in the present manuscript.\\ 
This paper is organized as follows. The photometric calibration of the clusters is described in Section~\ref{phca}. The existence of clusters is discussed in the Section~\ref{rdp_center}. The results of parameters through colour-magnitude diagrams (CMDs) and two colour diagrams (TCDs) are given in the Section~\ref{tcd_cmd}. Section \ref{mpa} is devoted to identify the probable members of IC \, 361 and its mean proper motions. Dynamical study of the cluster has been carried out in the Section~\ref{dyna}. The final conclusions and discussion are stated in Section \ref{co07}.

\begin{table*}
\caption{The observation details of $IC~361$. The value of observation time, exposure time and air-mass has been given in the same order.}
\begin{center}
\begin{tabular}{@{}|c|c|c|c|c|c|c|c|@{}}
\hline
Photomet- & No. of & Time of Observation & Exposure & Air-mass & Zero & Colour & Extinction\\
-ric Band  & frames & (JD=2456651+) & Time (Sec) & & point & Coefficient& Coefficient\\\hline
U & 1 & 0.27968 & 300 & 1.227 & $7.88{\pm}0.01$ & $-0.08{\pm}0.03$ & $0.26{\pm}0.01$ \\
B & 2 & 0.26235, 0.26404 & 30, 150 & 1.189, 1.194 & $5.33{\pm}0.01$ & $-0.06{\pm}0.02$ & $0.11{\pm}0.00$\\
V & 2 & 0.26810, 0.27126 & 120,30  & 1.200, 1.204 & $4.74{\pm}0.00$ & $-0.09{\pm}0.01$ & $0.07{\pm}0.00$\\
R & 1 & 0.27322 & 40 & 1.208 & $4.44{\pm}0.01$ & $-0.05{\pm}0.01$ & $0.04{\pm}0.00$ \\
I & 2 & 0.27513, 0.27690 & 25, 100 & 1.211, 1.216 & $4.91{\pm}0.01$ & $-0.06{\pm}0.01$ & $0.03{\pm}0.00$ \\
\hline
\end{tabular}
\end{center}
\label{tab1}
\end{table*}
\section{Photometric calibration and comparison with previous study}
\label{phca}
\begin{figure}[tbp]
\centering
\includegraphics[width=20pc,angle=0]{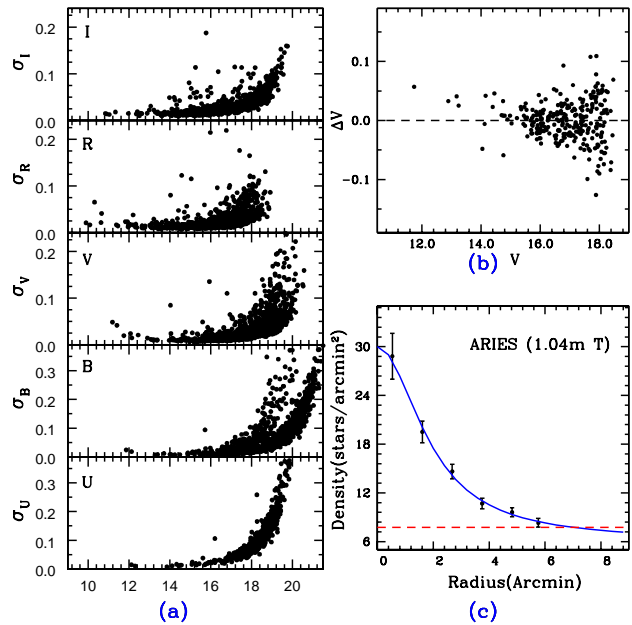}
\caption{\label{01_cmrd} {\bf a):-} The photometric error of stars as a function of brightness in the $UBVRI$ bands. {\bf b):-} A comparison of our $V$ observations with the $V$-magnitude of stars of work of Zdanavicius et al. 2010. The solid line represents a zero-magnitude difference. {\bf c):-} The stellar RDP of stars of the field of $IC~361$. The solid line represents the King profile and the horizontal line shows the average field star density.}
\end{figure}

Johnson-Cousins $UBVRI$ photometry of stars in the field of IC 361 was obtained on 2013, December 24 using the 1.04-m Sampurnanand telescope at Nainital, India. The observation detail of this cluster is given in the Table \ref{tab1}. On the same night we also observed Landolt's standard field: SA92 \citep{la92}. The usual image processing procedures (bias subtraction, flat fielding, and cosmic ray removal) were performed through IRAF\footnote{Image Reduction and Analysis Facility (IRAF) is distributed by the National Optical Astronomy Observatories, which are operated by the Association of Universities for Research in Astronomy, Inc., under cooperative agreement with the National Science Foundation.} software package. Photometry of the frames were performed using the DAOPHOT II profile fitting software \citep{st87}. The resultant transformation coefficients for the standard stars are also listed in the Table \ref{tab1} and photometric error with magnitude are depicted in Figure \ref{01_cmrd}a. We have found 1883 stars in the observed field of view (FOV) of the cluster.\\
In order to transform CCD pixel coordinates to celestial coordinates, we used the on-line digitized ESO catalogue included in the skycat software as an absolute astrometric  reference frame. A linear astrometric solution was derived for the V filter reference frame by matching positions of 170 well isolated bright stars. The $ccmap$ and $cctran$ routines are utilized for this purpose. We have been found 276 common stars between present photometry and Vilnious photometry \citep{zd10} and V-magnitude difference of both photometry is shown in the Figure \ref{01_cmrd}b.
\begin{figure}[tbp]
\centering
\includegraphics[width=20pc,angle=0]{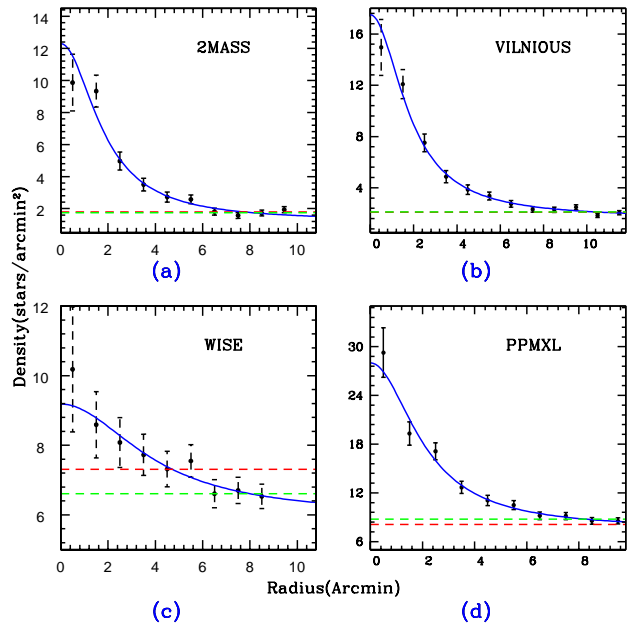}
\caption{\label{rdp} The radial density profiles of the cluster, IC 361 through various catalogue, namely {\bf a):-} 2MASS, {\bf b):-} VILNIOUS, {\bf c):-} WISE and {\bf d):-} PPMXL. The solid line of each panel represents the King profile and the red and green horizontal lines are representing the average field star density ($f_{0}+3\sigma$) and the average of first three points of field densities. The approximate equal value points are considered to be the points of field densities.}
\end{figure}
 \begin{table}
    \caption{Table represents the values of radius and core radius value of IC\,361 as estimated by various extracted/estimated photometric data-sets of catalogues/observatory.}
    \label{tab2_a}
    \medskip
    \begin{center}
      \begin{tabular}{l ccc} \hline
 Radius &  Core-radius & Max. Model & Av. Field\\
     (in arcmin)& (in arcmin) & Density &  Density\\\hline
      \end{tabular}\\[5pt]
      2MASS Catalogue\\
      \begin{tabular}{l ccc} \hline
      8.0 $\pm$ 0.4 & 1.83 $\pm$ 0.34 & 11.08 $\pm$ 2.15 & 1.21 $\pm$ 0.19\\
      \hline \end{tabular}\\[5pt]
      1.04-m telescope of ARIES\\
      \begin{tabular}{l ccc} \hline
     7.0 $\pm$ 0.2  & 1.91 $\pm$ 0.20 & 23.89 $\pm$ 1.76 & 6.11 $\pm$ 0.55\\
      \hline \end{tabular}\\[5pt]
      PPMXL Catalogue\\
      \begin{tabular}{l ccc} \hline
     8.2 $\pm$ 0.2 & 2.10 $\pm$ 0.24 & 20.48 $\pm$ 2.12 & 7.52 $\pm$ 0.29\\ 
      \hline \end{tabular}\\[5pt]
       WISE Catalogue\\
      \begin{tabular}{l ccc} \hline 
      8.0 $\pm$ 0.4 & 3.99 $\pm$ 1.42 & 3.21 $\pm$ 0.51 & 5.97 $\pm$ 0.45\\
      \hline \end{tabular}\\[5pt]
      Catalogue of VILNIOUS Observatory\\
      \begin{tabular}{l ccc} \hline 
      10.0${\pm}$ 0.4 & 1.85 $\pm$ 0.19 & 15.87 $\pm$ 1.74 & 1.63 $\pm$ 0.15\\
      \hline
      \end{tabular}\\[5pt]
    \end{center}
  \end{table}
%
\section{Center and Radius of the cluster}
\label{rdp_center}
The center coordinate (${\alpha},{\delta}$) of cluster is found to be $(4^h 18^m 56^s, +58^015'18'')$. Similarly, the equivalent pixel coordinate of center of observed reference of IC, 361 frame is given as (x,y)=(465, 545). The radial density profile (RDP) of open cluster is used to determine the cluster radius. The RDP is constructed by determining the stellar density in concentric rings of equal width around the cluster center. We choose a thickness of about 1 arcmin ($\sim$ 80 pixels) for each ring in order to have statistically significant number of stars in each radial zone. The stellar density ($\rho_{r}$ ) at the distance $r$ from the cluster center is defined by the following empirical formula \citep{kal92}:
\begin{equation}
\rho_{r} = \rho_{f} + \frac{\rho_{0}}{1+{(\frac{r}{r_c})}^2}
\end{equation}
where $\rho_{0}$ is the peak stellar density and $\rho_{f}$ is the background density. The $r_{c}$ is core radius of the cluster defined as a distance from the center at which the density $\rho_{r}$ becomes half of the $\rho_{0}$. The cluster radius is determined where RDP intercepts to background density. The prescribed background density is estimated as the average of first three approximate value of field densities and estimated background stellar densities of utilized data-sets of each catalogue have depicted by green line in each panel of Figure \ref{rdp}, whereas the model stellar densities have depicted by the red lines. The different seller number of different catalogue provide the different value of radius through the above said empirical formula. All obtained results are listed in the Table \ref{tab2_a}. To see obtained results, we conclude the size of radius and core radius of IC\, 361 as $8.24{\pm}0.32$ arcmin ($\sim$ 645 pixels) and $2.34{\pm}0.47$ arcmin, respectively.
\section{Analysis through the CMD and TCD}
\label{tcd_cmd}
\subsection{($U-B$) vs ($B-V$) TCD }\label{zams}
The reddening, $E(B-V)$, in the cluster region is estimated using the $(U-B)$ vs $(B-V)$ TCD. In Fig.~\ref{01_cmd}-b, we plot TCD for the members found in our study. We also draw zero age-main sequence [ZAMS \citep{sc82}] for observed MS stars on ($U-B$) vs ($B-V$) diagram. Here, we assume a solar metallicity for the cluster and a reddening vector of $E(U-B)/E(B-V)=0.72$ \citep{de78}. The  error in reddening is calculated using the following relation as given by \citep{ph94}.
\begin{equation}
\sigma_{E(B-V)_{sys}}^2 = \sigma^2_{(U-B)_{0}} + \sigma^2{(B-V)_{0}} + \sigma^2_{vector}
\end{equation}
The derived reddening E(B-V) from our study is $0.56\pm 0.10$, which is close to 0.55 estimated by \citet{zd10}. The colour-excess $E(V-R)$ and $E(V-I)$ are estimated by the relations $E(V-R)=0.60 \times E(B-V)$ and $E(V-I)=1.25 \times E(B-V)$ respectively. From
the relations, we estimated colour excesses of $E(V-R)$ and $E(V-I)$ as
$0.33{\pm}0.06$ and $0.7{\pm}0.12$ mag, respectively.

\subsection{Age and Distance by CMDs}\label{cmd}
For the determination of age and distance of the cluster, we draw the $(B-V)/V$ and $(V-I)/V$ CMDs for members of cluster as illustrated in Fig~\ref{01_cmd}-c which show a well populated but broad main sequence (MS) stars that may be due to photometric errors and/or presence of binary stars within the cluster. The variable reddening across the cluster region could also be the cause of the broad MS. We overplot Marigo's theoretical isochrones \citep{ma08} on the CMDs by varying the distance modulus and age simultaneously in both $(B-V)/V$ and $(V-I)/V$ CMDs while keeping reddening  $E(B-V)=0.60$ mag and $E(V-I)=0.75$ mag fixed as determined in the previous sub-section and assuming $E(V-I)=1.25 \times E(B-V)$. From the best visual isochrone fit to the varying age and distance combinations on the MS stars, we obtained a distance modulus $(V-M_V)=14.4\pm0.0.10$~mag and $log(Age)=9.10\pm0.05$~(yr) for the cluster IC\,361. Employing the correction for the reddening and assuming a normal reddening law (see, Sect.~\ref{s:ADS}), this corresponds to a true distance modulus $(m-M)_0 = 12.54\pm0.05$ mag or a distance of 3.22$\pm$0.07 kpc for the cluster. Our estimated distance is close to the distance of $3.3$ kpc obtained by \cite{zd10}.

Though isochrone fitting is often used to estimate the age of the cluster in the absence of more valuable but lesser available spectroscopic observations but it should be kept in mind that determining precise age through isochrone fitting in clusters, where no evolved stars are found, is very difficult as it contains a large uncertainty on this value.
\begin{figure}[tbp]
\centering
\includegraphics[width=20pc,angle=0]{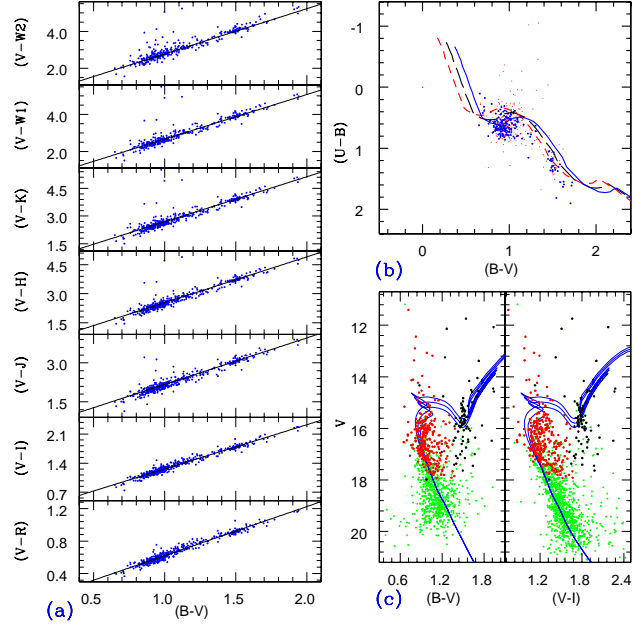}
\caption{\label{01_cmd} {\bf a):-} $(B-V)$ vs $(V-\lambda)$ diagrams, where $\lambda$ represents $R$, $I$, $J$, $H$, $K$, $W_1$, and $W_2$. {\bf b):-} $(B-V)$ vs $(U-B)$ colour-colour diagram for the cluster IC\,361. For clarity, stars having $eU>0.05$ mag are not shown. The arrow shows a slope of normal reddening vector $E(U-B)/E(B-V)$=0.72. The dashed black line show the best fit taken in account of 0.56 and 0.40 mag shift in $(B-V)$ and $(U-B)$, respectively while blue and red lines represent an error of $\pm$0.10 in the reddening vector. {\bf c):-} (B-V)/V and (V-I)/V CMDs in the field of cluster IC\,361. The solid line represents the best fit isochrone to the cluster for $\log$(Age)=$9.1{\pm}0.05$ and $(m-M)$=14.4 mag. The red and black dots represents the probable members and Giant members of IC\, 361, having V-magnitude greater than 18 Mag. Similarly, green dots represents the stars as found in the direction of field of view of the studied cluster IC\, 361.}
\end{figure}
\subsection{Nature of extinction law}
As the light emitted from the star cluster, it passes through the interstellar dust and gas, hence the scattering and absorption of light takes place. Since the absorption and scattering of the blue light is more compared to the red light, the stars appear reddened. Generally, normal reddening law is applicable when dust and intermediate stellar gases are absent in the line of sight of the cluster \citep{sn78}. However, the reddening law is expected to be different in the presence of dust and gas. We have investigated the nature of reddening law using $(V-\lambda)/(B-V)$ TCDs \citep{ch90}, where $\lambda$ is any broad band filter, namely $R$, $I$, $J$, $H$, $K$, $W_1$, and $W_2$. Here, $W_1$ and $W_2$ are mid-IR pass-bands  used for Wide-field IR Survey Explorer (WISE)  by \cite{wr10}. The $(V-\lambda)/(B-V)$ TCDs have been widely used to separate the influence of the extinction produced by the diffuse interstellar material from that of the intra-cluster medium \citep{ch90}. Therefore, we constructed the resultant $(V-\lambda)/(B-V)$ diagram in different wavelength bands, as shown in Fig.~\ref{01_cmd}-a. A best linear fit in the TCD of cluster gives the value of slope ($m_{cluster}$) for the corresponding TCD. The resultant values of the $m_{cluster}$ for seven colours are listed in Table~\ref{tab2} along with their normal values. Our slopes are quite comparable with those obtained for the diffuse interstellar material.

A total-to-selective extinction $R_{cluster}$ is determined using the relation given by \cite{ne81} as
\begin{equation}
R_{cluster}=\frac{m_{cluster}}{m_{normal}} \times R_{normal}.
\end{equation}
Assuming the value of $R_{normal}$ for the diffuse interstellar material as 3.1, we determined $R_{cluster}$ in first five colours and given in Table~\ref{tab2}. The mean value of $R_{cluster}$ is estimated as $R=2.94{\pm}0.11$ for the cluster IC\,361 through the last four computed values of $R_{cluster}$; moreover, this mean value of $R_{cluster}$ is low with compared to the normal reddening law. This prescribed Table indicates that the value $R_{cluster}$ is found to be high for the shorter effective wavelength with reference of $V$ optical band; whereas the values of $R_{cluster}$ are low for bands, having the high effective wavelength compare than the $V$ optical band.
  \begin{table}
    \caption{The slopes of the $(\lambda -V)/(B-V)$ TCDs in the direction of the cluster IC~361. Their normal values are also given in the second column.}
    \label{tab2}
    \medskip
    \begin{center}
      \begin{tabular}{r ccc} \hline
      Color & Normal & Estimated &  Total-to-Selective\\
      &Value & Value & Extinction Value \\\hline
      $\frac{R-V}{B-V}$ & 0.55 & 0.60 $\pm$ 0.01 & 3.38\\ \\
      $\frac{I-V}{B-V}$ & 1.10 & 1.06 $\pm$ 0.01 & 2.99\\  \\
      $\frac{J-V}{B-V}$ & 1.96 & 1.79 $\pm$ 0.05 & 2.83\\   \\
      $\frac{H-V}{B-V}$ & 2.42 & 2.37 $\pm$ 0.05 & 3.03\\  \\
      $\frac{K-V}{B-V}$ & 2.60 & 2.44 $\pm$ 0.05 & 2.91\\  \\
    $\frac{W_1-V}{B-V}$ & ---- & 2.45 $\pm$ 0.07 & ----\\  \\
    $\frac{W_2-V}{B-V}$ & ---- & 2.41 $\pm$ 0.07 & ----\\  \\ \hline
      \end{tabular}\\[5pt]
    \end{center}
  \end{table}
%
\subsection{Reddening in near-infrared bands}\label{s:ADS}The two-colour diagrams (TCDs) plots of various sets of two colours (magnitude difference of two bands) are important tools for investigating the nature of extinction law, dependency with each other and their variation from normal values (Joshi {\&} Tyagi 2016). We are prescribing about two TCDs as below,
\subsubsection{(V-K) vs (J-K) TCD}
\begin{figure}[tbp]
\centering
\includegraphics[width=20pc,angle=0]{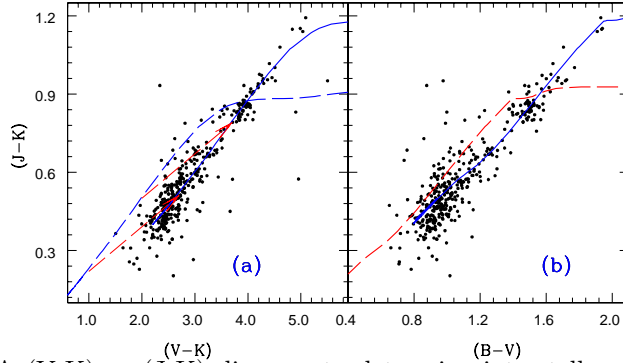}
\vspace{-4cm}
\caption{\label{infra} {\bf a):-} A (V-K) vs (J-K) diagram to determine interstellar extinction is shown in the left panel. The blue dashed line represents Marigo’s isochrones of Solar metallacity without any shift in colour while blue solid line represents best fitted line after shift of 1.72 mag in x-direction and 0.29 mag in y-direction. Two red arrows represent the reddening vector $\frac{(J-K)}{V-K}=0.173$. {\bf b):-} A (B-V) vs (J-K) diagram is depicted in the right panel.}
\end{figure}

A $(V-K)$ vs $(J-K)$ diagram is used to determine the interstellar extinction in the near-IR (NIR) range. In Fig.~\ref{infra}-a, we plot $(V-K)$ vs $(J-K)$ diagram where we also overplot Marigo isochrones of solar metallicity by shifting the line in the direction of reddening vector $\frac{E(J-K)}{E(V-K)}=0.173$. In this way, we obtained colour excesses of $E(V-K)=1.72{\pm}0.05$ mag and $E(J-K)=0.29{\pm}0.03$ mag. The relation $R=1.1 \times \frac{E(V-K)}{E(B-V)}$ given by \cite{wh80} was used to estimate the corresponding value of reddening. In Fig.~\ref{infra}-a, we draw reddening vectors $\frac{E(V-K)}{E(B-V)}$ by shifting for the reddening values in the NIR bands (continuous line) and optical bands (long-dashed line). We yield $E(B-V)=0.62{\pm}0.03$ mag in the direction of the cluster which is slightly higher than that was determined in the optical band. It clearly suggests a small IR excess for stars in the cluster IC~361. However, excess emission in the cluster IC~361 is not consistent with many relatively young clusters. Thus, IC~361 is an old age cluster with low IR-excess. 
\subsubsection{(B-V) vs (J-K) TCD}
This prescribed TCD for present studied cluster is shown in the right panel of Fig. \ref{infra}. The Marigo isochrone is used to know relationship between E(B-V) and E(J-K). The colour excess E(B-V) and E(J-K) of IC 361 are found to be 0.56 and 0.30 mag respectively. As a result, the E(J-K)/E(B-V) ratio for IC 361 is computed to be
0.54, which is close to the literature value, i.e. 0.56 \citep{joshi16}. This variation is possible due to the different value of total-to-selective-extinction from normal value. The red line of Fig \ref{infra}-b is representing Marigos isochrone without any shift, whereas the blue line is showing the best fitted line after shift of 0.56 mag in x-direction and 0.30 mag in y-direction respectively.
\section{Mean Proper Motion and Probable members}\label{mpa}
\begin{figure}[tbp]
\centering
\includegraphics[width=20pc,angle=0]{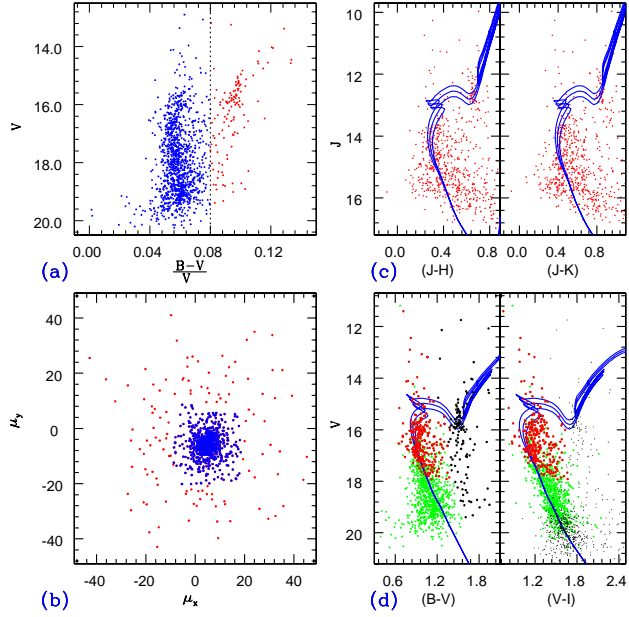}
\caption{\label{01_cmd} {\bf a):-} The CMRD of IC\, 361 is shown in upper left-hand panel. The red dots are depicted to the stellar branch of Giant Stars. {\bf b):-} The proper-motion distribution of stars for IC\, 361. The red dots represent the statistically cleaned probable members used to determine the mean-proper motion while the blue dots represent the probable members used to determine the mean proper motion of the cluster. {\bf c):-} (J-H)/J and (J-K)/J CMDs in the field of cluster IC\,361. The blue solid line represents the best fit isochrone to the cluster for $\log$(Age)= $9.1{\pm}0.05$ and $(m-M)_{J}$=12.87$\pm$0.04 mag. {\bf d):-} (B-V)/V and (V-I)/V CMDs in the field of cluster IC\,361. The solid line represents the best fit isochrone through $\log$(Age)=$9.1{\pm}0.05$ and $(m-M)_{V}$=14.4 mag.}
\label{fig04}
\end{figure}

The mean proper-motion value of the cluster is the mean value of proper-motion values of its probable members. Probables members of IC 361 are found through the field star decontamination. For this purpose, we have applied CMRD \citep{joshi16} approach to find clear stellar sequence of studied cluster. The CMR value of $\frac{B-V}{V}$ is 0.08 for brighter stars of cluster and this value is used to separate cluster's stellar sequence from the field sequence as depicted in Figure~\ref{fig04} (a). The cluster radius is found to be 8 arcmin in present analysis leads the value of cluster's diameter is $16{\pm}1$ arcmin. Since, estimated diameter is greater than the size of observed optical frame as well as other available optical catalogues, therefore, we can not utilized these catalogues for statistical colour-magnitude cleaning \citep{joshi15} due to observations of limited region of cluster. The available $2MASS$ data of field region of IC 361 is fulfilled criteria of our need. As a result, NIR $JHK$ data of cluster is used to further decontamination of filed stars. Since, required area of selected field region is equivalent to the enclosed area of cluster periphery, therefore, the disc size of field region for IC 361 is taken to be $10.0~arcmin~{\leq}~r~{\leq}~12.8~arcmin$.\\ 
After iterating the cleaning procedure for each field region star, we found 850 probable members on the statistically cleaned CMDs (detail of cleaning procedure is published by Joshi et al., 2015). The location of these probable members around best fitted isochrones of $(B-V)/V$ and $(J-K)/J$ CMDs confirms the main sequence of IC 361. The resultant probable members are further utilized to estimate the mean proper motion of IC 361 through 3${\sigma}$ clipping approach. In this approach, the stars which do not fall within 3$\sigma$ value of the mean are rejected for estimation. After 3${\sigma}$ clipping, the mean proper motion values of IC 361 are found to be $\bar{\mu_{x}} = 4.97 {\pm}0.17~mas~yr^{-1}$ and $\bar{\mu_{y}} = -5.80 {\pm}0.18~mas~yr^{-1}$ in RA and DEC directions, respectively.
%
\section{Dynamical study of the cluster}\label{dyna}
\subsection{Luminosity and Mass functions} \label{lf}
The luminosity function (LF) is the total number of cluster members in different magnitude bins. The estimated number of stars in each magnitude bin for the cluster $(N_C)$ are given in Table~\ref{tab3}. One can see that number of stars reduced significantly after the 19 mag which is due to incompleteness factor. Here, we note that generally data incompleteness increases with the increasing magnitude and in the present catalogue, it is better than 90\% upto 19 mag \citep{sh08}. It is noted fact that the red dots are not appeared after 18 V-mag (See Figure 3-c), wherease the green dots are found uniformly with red dots and also appear after 18 V-mag. The turn point of best fitted isochrone finds below 15 V-mag and . We, therefore, have not applied any correction to the luminosity and mass functions between V-mag range 15-18 derived in the present study.
\begin{table}
\caption{The MF of the cluster IC 361}
\label{tab3}
\tiny
\begin{center}
\begin{tabular}{@{}ccccccc@{}}
\hline\hline
V range & Mass range & $\bar{m}$ & $log(\bar{m})$ & $N$  & $log(\Phi)$ & $e_{log(\Phi)}$ \\\\
(mag) & $M_{\odot}$  & $M_{\odot}$ & & & & \\
\hline%
13-14  & 2.160-2.345 & 2.252 & 0.353 & 045 & 3.103 & 0.149\\
14-15  & 2.149-2.160 & 2.155 & 0.333 & 110 & 4.696 & 0.095\\
15-16  & 1.878-2.149 & 2.014 & 0.304 & 203 & 3.540 & 0.070\\
16-17  & 1.573-1.878 & 1.726 & 0.237 & 305 & 3.598 & 0.057\\
17-18  & 1.312-1.573 & 1.443 & 0.159 & 386 & 3.690 & 0.051\\
18-19  & 1.104-1.312 & 1.208 & 0.082 & 283 & 3.578 & 0.059\\  
\hline
\end{tabular}
\end{center}
\end{table}
\begin{figure}
\begin{center}
\includegraphics[width=13.0pc, angle=270]{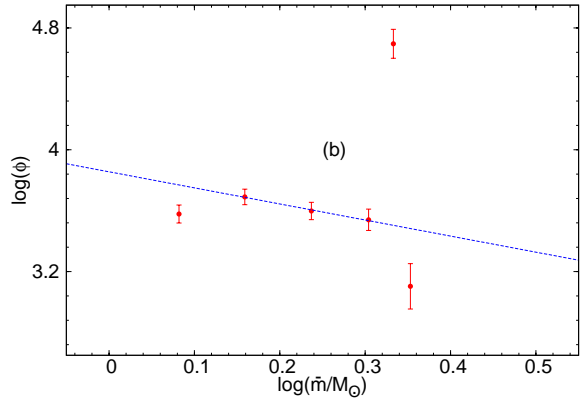}
\caption{The variation of logarithm MF with the logarirhm of average mass of members.}\label{fig5}
\end{center}
\end{figure}
The initial mass function (IMF) is defined as the distribution of stellar masses per unit volume in a star formation event and knowledge of IMF is very effective to determine the subsequent evolution of cluster. The direct measurement of IMF is not possible due to the dynamical evolution of stellar systems though we can estimate the present mass function (MF) of cluster. The MF is defined as the relative numbers of stars per unit mass and can be expressed by a power law $N(\log M) \propto M^{\Gamma}$. The slope, $\Gamma$, of the MF can be determined from
\begin{equation}
\Gamma = \frac{d\log N(\log\it{m})}{d\log\it{m}}
\end{equation}
where $N\log(m)$ is the number of stars per unit logarithmic mass. The masses of probable members can be determined by comparing observed magnitudes with those predicted by a stellar evolutionary model if the age, reddening, distance and metallicity are known. The MF determined for the cluster is given in Table~\ref{tab3}. The effect of field star contamination is negligible for estimation of MF due to fact that MPMs are used for this purpose in the present study. Fig.~\ref{fig5} shows the MF in the cluster fitted for the MS stars with masses $1.104{\leq}M/M_{\odot}~<~2.345$. The error bars have computed assuming Poisson Statistics. The MF slope ($\Gamma$) for cluster comes out to be $-1.06{\pm}0.09$ which is low compare to Salpeter MF slope of -2.35 \citep{sal} within the given uncertainty.
%
\subsection{Mass Segregation}
\begin{figure}
\includegraphics[width=20pc, angle=0]{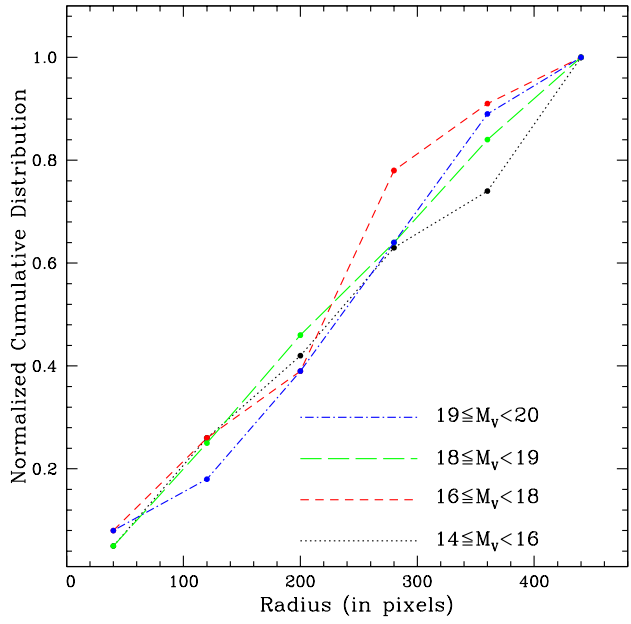}
\caption{The cumulative radial distribution of stars in various mass ranges of IC 361.}
\label{fig7}
\end{figure}

\begin{figure}
\includegraphics[width=20pc, angle=0]{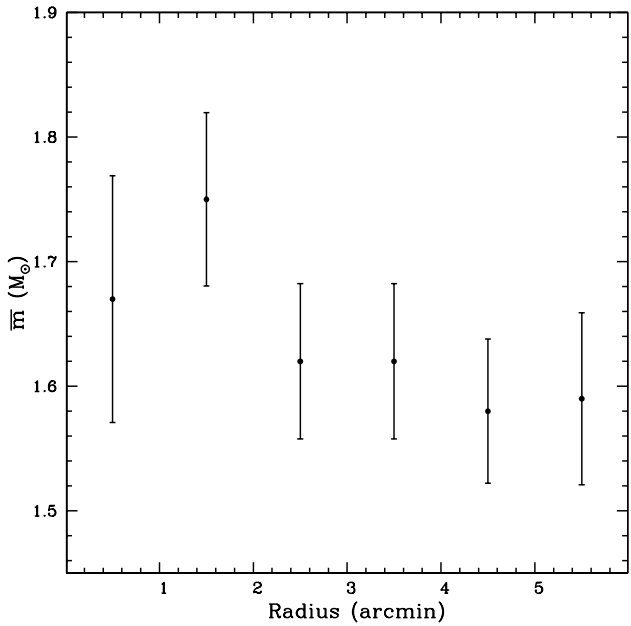}
\caption{The Mass distribution of stars in the radial direction.}
\label{fig6}
\end{figure}

Through the process of mass segregation, the stellar encounters of the cluster members are gradually increased. In the processes of mass segregation, the higher-mass cluster members gradually sink towards the cluster center by transferring their kinetic energy to the more numerous lower-mass stellar members of the cluster (Mathieu \& Lathen, 1986). With the dynamical evolution of the cluster, the spatial mass distribution in the cluster changes with time. To investigate the dynamical evolution and mass segregation process in the cluster IC~361, the cumulative radial stellar distribution of the probable members for various mass ranges are shown in the Figure~\ref{fig7} and also given in the Table \ref{tab3_a}. It is seen that MS slope does not become steeper as the radial distance from the cluster increases. This clearly suggests that the mass segregation is not taking place in the cluster. To further study the mass segregation in more detail, we plot the variation of mean mass along the radial distance for the probable members in Fig.~\ref{fig6}. The average mass of center is low as expected but satisfy the condition of stellar enhancement. 
\begin{table}
    \caption{Table represents the estimated normalized cumulative distribution of probable members of IC\, 361 in various stellar magnitude groups.}
    \label{tab3_a}
    \medskip
    \begin{center}
      \begin{tabular}{l ccc} \hline
 Zone Size &  Number & Cum. & Norm. cum. \\
     (in pixels)& of stars & stellar No. &  Distribution\\\hline
      \end{tabular}\\[5pt]
      14 ${\leq}M_V < $ 16\\
      \begin{tabular}{l ccc} \hline
      000-080 & ~~~~   001 ~~~~   &  ~~~~  001  ~~~~   &  ~~~~  0.05  ~~~~  \\
      080-160 & ~~~~   004 ~~~~   &  ~~~~  005  ~~~~   &  ~~~~  0.26  ~~~~  \\
      160-240 & ~~~~   003 ~~~~   &  ~~~~  008  ~~~~   &  ~~~~  0.42  ~~~~  \\
      240-320 & ~~~~   004 ~~~~   &  ~~~~  012  ~~~~   &  ~~~~  0.63  ~~~~  \\
      320-400 & ~~~~   002 ~~~~   &  ~~~~  014  ~~~~   &  ~~~~  0.72  ~~~~  \\   
      400-480 & ~~~~   005 ~~~~   &  ~~~~  019  ~~~~   &  ~~~~  1.00  ~~~~  \\
      
      \hline \end{tabular}\\[5pt]
      16 ${\leq}M_V < $ 18\\
      \begin{tabular}{l ccc} \hline
      000-080 & ~~~~   029 ~~~~   &  ~~~~  029  ~~~~   &  ~~~~  0.08  ~~~~  \\
      080-160 & ~~~~   071 ~~~~   &  ~~~~  100  ~~~~   &  ~~~~  0.26  ~~~~  \\
      160-240 & ~~~~   047 ~~~~   &  ~~~~  147  ~~~~   &  ~~~~  0.39  ~~~~  \\
      240-320 & ~~~~   048 ~~~~   &  ~~~~  295  ~~~~   &  ~~~~  0.78  ~~~~  \\
      320-400 & ~~~~   051 ~~~~   &  ~~~~  346  ~~~~   &  ~~~~  0.91  ~~~~  \\  
      400-480 & ~~~~   032 ~~~~   &  ~~~~  378  ~~~~   &  ~~~~  1.00  ~~~~  \\
      \hline \end{tabular}\\[5pt]
      18 ${\leq}M_V < $ 19 Catalogue\\
      \begin{tabular}{l ccc} \hline
      000-080 & ~~~~   012 ~~~~   &  ~~~~  012  ~~~~   &  ~~~~  0.05  ~~~~  \\
      080-160 & ~~~~   042 ~~~~   &  ~~~~  054  ~~~~   &  ~~~~  0.25  ~~~~  \\
      160-240 & ~~~~   047 ~~~~   &  ~~~~  101  ~~~~   &  ~~~~  0.46  ~~~~  \\
      240-320 & ~~~~   040 ~~~~   &  ~~~~  141  ~~~~   &  ~~~~  0.64  ~~~~  \\
      320-400 & ~~~~   042 ~~~~   &  ~~~~  183  ~~~~   &  ~~~~  0.84  ~~~~  \\  
      400-480 & ~~~~   036 ~~~~   &  ~~~~  219  ~~~~   &  ~~~~  1.00  ~~~~  \\
      \hline \end{tabular}\\[5pt]
       19 ${\leq}M_V < $ 20\\
      \begin{tabular}{l ccc} \hline 
      000-080 & ~~~~   019 ~~~~   &  ~~~~  019  ~~~~   &  ~~~~  0.08  ~~~~  \\
      080-160 & ~~~~   026 ~~~~   &  ~~~~  045  ~~~~   &  ~~~~  0.18  ~~~~  \\
      160-240 & ~~~~   052 ~~~~   &  ~~~~  097  ~~~~   &  ~~~~  0.39  ~~~~  \\
      240-320 & ~~~~   060 ~~~~   &  ~~~~  157  ~~~~   &  ~~~~  0.64  ~~~~  \\
      320-400 & ~~~~   063 ~~~~   &  ~~~~  220  ~~~~   &  ~~~~  0.89  ~~~~  \\  
      400-480 & ~~~~   027 ~~~~   &  ~~~~  247  ~~~~   &  ~~~~  1.00  ~~~~  \\
      \hline \end{tabular}\\[5pt]
    \end{center}
  \end{table}
%
Our study indicate that the higher and lower mass stars are uniformly distributed in the cluster region. To understand whether it is an imprint of star formation process in the clusters and/or result of dynamical evolution, we determined dynamical relaxation time, $T_E$, which is the time in which individual stars in the cluster exchange energies and their velocity distribution approaches Maxwellian equilibrium. It can be expressed as
\begin{equation}
T_E = \frac{8.9 \times 10^5 (N R_h^3/\bar{m})^{1/2}} {\log(0.4N)}
\end{equation}
where $T_E$ is in Myr, $N$ is the total number of cluster members, $R_h$ is the radius (in parsecs) containing half of the cluster mass and $\bar{m}$ is mean mass of the cluster members in solar units (cf. Spitzer \& Hart, 1971). We estimated a total of 1362 member stars in the mass range $0.93 \le M/M_\odot < 2.4$. The total mass of the cluster is found to be $\sim 2198.95 M_\odot$ for IC\,361, which gives an average mass of $\sim 1.61 M_\odot$ per star. The contribution of the low-mass stellar population is critical for constraining the total cluster mass, which is crucial in understanding the dynamical evolution and the long-term survival of a cluster (e.g., de Grijs \& Parmentier 2007, and references therein). It can be seen that the half-radius of the cluster, $R_h$, plays an important role in the determination of the dynamical relaxation time, $T_E$. We obtained a half-mass radius for the cluster as 1.00 pc which is $\sim 54.8\%$ of the cluster radius. A  half-mass radius larger than half of the cluster radius suggests that inner region still has a deficiency of massive stars in the core region of the cluster. We estimated the dynamical relaxation time $T_E$ = 9.46\,Myr for the cluster. $T_E$ determined in the present study is very lower than the present age of about 1258.92\,Myr which suggests that cluster IC~361 is dynamically relaxed.
\section{Discussion and Conclusion}
\label{co07}
Open clusters are important tools for the understanding of the evolution phenomenon in galaxies because they provide information on star formation history. In the present study, we have constructed a complete $UBVRIJHKW_1W_2$ catalogue for the cluster IC 361 by supplementing our photometric data with the 2MASS $NIR$ and WISE mid-$IR$ surveys. Using a $V$ -band data of the present catalogue for the stars brighter than $V$ = 18 mag, we found that the core and cluster radius are $2.0 {\pm} 0.4$ arcmin  and $8.0 {\pm} 0.5$ arcmin, respectively. Using probable members, we derived the reddening in optical band using $(U - B)/(B - V )$ diagram and in NIR band using $(V - K)/(J - K)$ diagram. The values of reddening were obtained as $0.56{\pm}0.10$ mag and $0.62{\pm}0.03$ mag in optical and NIR, respectively. Since IC 361 shows a differential extinction across the cluster region as well as IR colour excess, it indicates that the stars are still embedded in the parent molecular gas and dust.\\
Based on the $(B -V )/V$ and $(V -I)/V$ CMDs and a visual fitting of isochrones for solar metallicity given by Marigo et al. (2008) to the blue sequence on the CMD, we determined a distance of $3.22{\pm}0.07$ kpc and log(age) of $9.10 {\pm} 0.05$ for the cluster IC 361. At this distance, we estimated a respective core and cluster radii of $7.50 {\pm} 0.94$ pc and $1.87 {\pm} 0.18$ pc for this cluster. We estimated the average total-to- selective extinction value as $2.94{\pm}0.11$ that is low compare to the normal value. The mean proper motion of the cluster was determined using the PPMXL catalogue  and found to be $4.97 \pm 0.17 ~mas/yr$ and $-5.80 \pm 0.18 ~mas/yr$ in the direction of RA and DEC, respectively. The MF slope of probable stars of the cluster is found to be $−1.06 {\pm} 0.09$ in the mass range $1.10{\leq}M/M_{\odot} < 2.34$. We do not have any evidence of stepper MF slope with the radial distance which suggests that the mass segregation does not take place in the cluster. The relaxation time of the cluster has found to be much lower than its age which implies that the cluster has yet dynamically relaxed.
\section*{Acknowdege}
GCJ is acknowledged to ARIES, Nainital for providing observing facility duration Oct, 2012 to April 2015. GCJ is thankful to Dr. A.K. Pandey, Director, ARIES (Nainital) to permit to him to use to gathered raw data of candidate through the letter No. AO/2018/41 on date 12 April 2018. This publication made use of data products from the Two Micron All Sky Survey, which is a joint project of the University of Massachusetts and the Infrared Processing and Analysis Center/California Institute of Technology, funded by the National Aeronautics and Space Administration and the National Science Foundation. This publication is also makes use of data products from the Wide-field Infrared Survey Explorer, which is a joint project of the University of California, Los Angeles, and the Jet Propulsion Laboratory/California Institute of Technology, funded by the National Aeronautics and Space Administration.


\bibliographystyle{model1a-num-names}
\bibliography{<your-bib-database>}

\end{document}